# Evidence for existence of Functional Monoclinic Phase in Sodium Niobate based Solid Solution by Powder Neutron Diffraction

S. K. Mishra[1], Mrinal Jauhari[1,2], R. Mittal[1,2], P. S. R. Krishna[1] V. R. Reddy[3] and S. L. Chaplot[1,2]

[1]*Solid State Physics Division, Bhabha Atomic Research Centre, Trombay, Mumbai- 400085, India-*

[2]*Homi Bhabha National Institute, Anushaktinagar, Mumbai 400094, India*

[3]*UGC-DAE Consortium for Scientific Research, University Campus, Khandwa Road, Indore 452001, India*

## Abstract

We have carried out systematic temperature-dependent neutron diffraction measurements in conjunction with dielectric spectroscopy from 6 to 300 K for sodium niobate based compounds $(1-x)$ $NaNbO_3$ - $xBaTiO_3$ (NNBTx). The dielectric constant is measured both as a function of temperature and frequency. It shows an anomaly at different temperatures in cooling and heating cycles and exhibits a large thermal hysteresis of ~150 K for the composition x=0.03. The dielectric constant is found to be dispersive in nature and suggests a relaxor ferroelectric behavior. In order to explore structural changes as a function of temperature, we analyzed the powder neutron diffraction data for the composition x=0.03 and 0.05, respectively. Drastic changes are observed in the powder profiles near $2\theta \sim 30.6°$, $32.1°$ and $34.6°$ in the diffraction pattern below 200 K during cooling and above 190 K in heating cycles, respectively. The disappearance of superlattice reflection and splitting in main perovskite peaks provide a signature for structural phase transition. We observed stabilization of a monoclinic phase (Cc) at low temperature. This monoclinic phase is believed to provide a flexible polarization rotation and considered to be directly linked to the high performance piezoelectricity in materials. The thermal hysteresis for composition x=0.03 is larger than x=0.05. This suggests that addition of $BaTiO_3$ in $NaNbO_3$ suppresses the thermal hysteresis. It is also observed that the structural phase transition temperature decreases on increasing dopant concentration.

Sodium Niobate and its solid solution has received attention as a potential candidate for replacement of commonly used lead based piezoelectric materials[1-20]. Sodium Niobate is antiferroelectric at room temperature and shows a characteristic double hysteresis loop in high quality single crystal during initial cycles only. However, a characteristic usually associated with ferroelectrics, square polarization hysteresis loops are observed in polycrystalline $NaNbO_3$. It is important to state that it is actually antiferroelectric at room temperature in the as-sintered state(virgin). Application of a small critical electric field in $NaNbO_3$ transforms it into a metastable ferroelectric phase[21]. Low temperature neutron diffraction study of sodium niobate also provides unambiguous evidence for the coexistence of ferroelectric and antiferroelectric phases[22]. These experimental observations suggest that there is small difference between the free energies of antiferroelectric and ferroelectric phases and further confirmed by theoretical studies also[22, 23]. The small difference between ferroelectric and antiferroelectric phases makes sodium niobate a suitable candidate to tailor its crystal structure and properties easily by suitable change in the thermodynamical conditions and application of external or chemical pressure. For example, substitution of Li or K for Na induces a ferroelectric phase that provides piezoelectric activity, which is interesting for technological application[24-28].

To explore the eco-friendly piezo-ceramics, recent studies are focused on the analysis, involving structural characterization and dielectric measurements, of the phase transitions of $NaNbO_3$-$ABO_3$ [3, 16, 18-20, 23-35] systems. Substitution of different chemical species with similar ionic charge like Li and K can be treated as an equivalent to pressure and properties may be rationalized with respect to end members. However, substitution with different ionic charges has a two-fold effect: (i) create a chemical pressure due to difference in their ionic radii and (2) charge distribution result of difference of their charges. The formation of long-range polar order is prohibited by the random field and random bonds in solid solution having multiple cations with differences in size, valency, or polarizability occupying crystallographically equivalent sites. As a result of this short-range ordered relaxor (R) state is expected to form. Relaxor-based solid solutions are used in a wide variety of piezoelectric applications owing to their excellent electromechanical properties[36]. Substitution of $CaTiO_3$ and $SrTiO_3$ (known as incipient ferroelectrics) in $NaNbO_3$ strongly modified the crystal structure and enhanced the dielectric constant along with a morphotropic phase boundary.

Recently, the solid solutions (1-x) $NaNbO_3$- $xBaTiO_3$ (NNBTx) have been widely investigated by different workers. Zuo et.al[18, 19] has reported giant electrostrictive effect in the NNBTx relaxor ferroelectrics for the compositions x=0.15-0.25. So far, NNBTx solid solutions have been studied with main focus on their dielectric properties and relaxor behaviour. The doping of small amount of $BaTiO_3$ is expected to distort the sodium niobate lattice and show variety of structural phase transitions as a function of composition and temperature. To address this issue, we have carried out detailed studies of the compounds using a combination of dielectric and X-ray diffraction measurements[20]. We observed anomalies in dielectric measurements performed for various compositions at 300 K and found appearance/disappearance of the superlattice reflections along with change in the intensities of main perovskite peaks in the powder X-ray diffraction data. These changes provide clear evidence for structural phase transitions as a function of composition. We found that on increasing the concentration of $BaTiO_3$, the value of the polar mode (GM4-) first decreases up to x=0.05 and then increases. On the other hand, the value of one of the antiferrodistortive modes (M3+) increases up to x=0.07 and then decreases with concentration. However, the value of the antiferrodistortive (R4+) mode decreases sharply and vanishes for x> 0.07. We provided evidence that suppression of in-phase rotation of octahedra and increment in tetragonality (c/a ratio) promote the polar mode at room temperature. The temperature dependent powder X-ray diffraction study indicated that the ferroelectric rhombohedral phase of pure sodium niobate gets suppressed for the composition x=0.03, and the monoclinic phase *Cc* gets stabilized at low temperature. These phase transitions are associated with the position of oxygen. Neutron diffraction offers certain advantages over X-rays especially in the accurate determination of the oxygen positions.

In the present study, we have carried out systematic temperature-dependent neutron diffraction measurements in conjunction with dielectric spectroscopy of NNBTx with x=0.03 and 0.05 from 6 to 300 K. The temperature dependent dielectric constant measurement shows a dispersion with frequency, which suggests a relaxor ferroelectric behavior. It also exhibits anomalies at different temperatures in cooling and heating cycles and exhibits a large thermal hysteresis of ~150 K. This provides a signature for the first order nature of phase transition across this temperature. In order to explore if there is any structural change as a function of temperature, we analyzed the powder neutron diffraction data. We observed stabilization of monoclinic phase (Cc) which is identical to PZT where piezo response is ultra large. This monoclinic phase is believed to provide a flexible polarization rotation and also considered to be directly linked to the high performance piezoelectricity in materials.

Samples of NNBTx with x=0.03 and 0.05 were prepared by solid state reaction method[8]. The neutron powder diffraction data have been recorded in the 2θ range of 4°–138° a step of width of 0.1° using neutrons of wavelength of 1.2443 Å on a medium resolution position sensitive detector based powder diffractometer at the Dhruva Reactor in Bhabha Atomic Research Centre. The structural refinements were performed using the Rietveld refinement program FULLPROF[37] and details are given in supplementary material.

Figure 1 show the variation of dielectric constant as a function of temperature for NNBT03 at selected frequencies namely 1, 10 and 100 K Hz. It is evident from this figure that on cooling from room temperature, the magnitude of dielectric permittivity value decreases from 300 (at 300K) to 70 (at 5 K) with an anomaly around 160 K. Further, on heating from 5 K, it increases and show anomaly around ~ 250 K. The dielectric data show dispersion with frequency. The observed weak relaxor behaviour in NNBTx is due to the competing structural instabilities, rather than due to the structural heterogeneities as in the classical relaxor $Pb(Mg_{1/3}Nb_{2/3})O_3$ (PMN)[38]. It is also important to notice that these anomalies occur at different temperatures in cooling and heating cycles and exhibit a large thermal hysteresis of ~150 K. This provides a signature for the first-order nature of phase transition across this temperature. In order to explore if there is any structural change as a function of temperature, we have performed powder neutron diffraction experiments.

Figure 2 depicts a portion of the powder neutron diffraction patterns of (1-x) $NaNbO_3$ -x$BaTiO_3$ for x=0.03 and 0.05, at selected temperatures in the range of 6–300 K for cooling and heating cycles. The powder diffraction pattern consists of two types of reflections, viz., main peaks of perovskite structure (as seen in cubic phase) and additional reflections known as superlattice reflections. These superlattice reflections are a result of the octahedral tilting and are associated with zone boundary instability. The main peaks of perovskite structure can be noticed at around 2θ ~ $18.3^0$, $25.8^0$, $32.1^0$, $37.1^0$ and $41.8^0$ and superlattice peaks appear around 2θ ~ $28.9^0$, $30.6^0$, $34.6^0$, $37.1^0$ and $40.8^0$ at room temperature. It is evident from Fig. 2 that there is a drastic change in the profile near 2θ ~ 30.6°, 32.1° and 34.6° in the diffraction pattern below 200 K during cooling and above 190 K in heating cycles, respectively. As we lowered the temperature from 300 K, the intensity of the superlattice reflections at around 2θ ~ $28.9^0$ and $34.6^0$ decreases while the intensity of peak around 2θ ~$30.6^0$ increases. We also see a splitting in the main perovskite peak around 32.1 degree. The disappearance of superlattice reflection and splitting in main peak provide a signature for structural phase transition in both compounds.

One of the end-member, sodium niobate, exhibits a very complex sequence of structural phase transitions. It undergoes a series of antiferrodistortive phase transitions starting from paraelectric cubic (***Pm-3m***) to a ferroelectric rhombohedral ***R3c*** structure via intermediate paraelectric tetragonal and orthorhombic phases and an antiferroelectric orthorhombic phase[22]. It has an antiferroelectric phase with space group ***Pbcm*** at room temperature 300 K. The antiferroelectric phase consists of two or more sublattice polarizations of antiparallel nature, which in turn give rise to superlattice reflections in the diffraction patterns. We have observed such reflections in the antiferroelectric phase, and the strongest one of them appears at 2θ~ 29.5°. These characteristic antiferroelectric reflections are not present in powder neutron diffraction pattern of NNBT03 at room temperature, which suggests that it has a different crystal structure at 300 K.

The doping of small amount of ions such as $K^+$, $Li^+$ in pure $NaNbO_3$ imposes chemical pressure which results in the deviation of the crystal structure from its current antiferroelectric phase and stabilization of ferroelectric orthorhombic phase with space group ***$Pmc2_1$*** and cell dimensions 2×√2×√2 (with respect to elementary cubic perovskite cell). We refined powder neutron diffraction using this phase and found that all the peaks in the powder neutron diffraction patterns at 300 K could be indexed. The fit between the observed and calculated profiles is found to be satisfactory (Figure 3 (a)). The refined structural parameters are given in the table S1 (supplementary materials). This structure is identified as ferroelectric in nature by its symmetry. To confirm the ferroelectric nature of system, we carried out PE hysteresis loop at room temperature for composition x=0.03 and shown in figure 3 (b). It is evident from this figure that composition x=0.03 has ferroelectric loop with applied voltage. This confirm the ferroelectric nature for the samples. We have also carried out PE hysteresis loop at selected low temperatures and the results are shown in Figure 3b. It is evident from this figure that on decreasing temperature, the spontaneous polarization decreases, which is consistent with changes in the crystal structure.

Detailed Rietveld refinement of the powder-diffraction data shows that diffraction patterns could be indexed using the orthorhombic structure (space group *$Pmc2_1$*) up to 150 K in cooling cycle. The Rietveld refinements proceeded smoothly, revealing a monotonic decrease in lattice constant and cell volume with decreasing temperature. However, attempts to employ the same orthorhombic structural model in the refinements proved unsatisfactory at lower temperatures, and a progressive worsening of the quality of the Rietveld fits with decreasing temperature was found. The most apparent signature of the subtle structural transformation that occurs below 150 K is the inability of orthorhombic structure to account satisfactory for the peaks around 30.8°.

Recently, we have employed systematic neutron diffraction measurements as a function of temperature to study the low temperature structures and to understand the phase transitions of $NaNbO_3$ and its solid solution[25, 26] $(Na/Li) NbO_3$. Our studies present unambiguous evidence for the presence of the ferroelectric R3c phase at low temperature. Thus, we refined powder neutron diffraction data at 6 K using rhombohedral R3c phase (see inset in figure 4). The unsatisfactory quality of the Rietveld fit of the diffraction data suggests that the possibility of rhombohedral phase is not favoured in this compound. We tried to refine the powder neutron diffraction data using different crystal structure of classical perovskite ($PbTiO_3$ and $BaTiO_3$) exhibit at different condition but none of these space groups accounted for the Braggs peaks marked with arrow and account the splitting in peaks appeared 32.2 degree. We have further explored various possibilities as in our x-ray diffraction study and we found that monoclinic phase with space group Cc accounted nearly all the reflections at lowest temperature. In the space group Cc, for the description of the crystal structure, we used the following monoclinic axes: $a_m = a_p + b_p + 2c_p$; $b_m = a_p - b_p$ and $c_m = a_p + b_p$. The monoclinic unit cell contains 4 formula units. The asymmetric unit of the structure consists of Na/Ba atom at the 4a site at (o±u, ¾ ±v, 0±w), Nb/Ti atom at the 4a site at (¼±u, ¼±v, ¼±w). Three oxygen atoms occupies at the 4a site at (0±u, ¼±v, 0±w); (¼±u, ½ ±v, 0±w) and (¼±u, 0±v, ½±w) respectively. The fit between the observed and calculated profiles is quite satisfactory and include the weak superlattice reflections. However, we find some discrepancy in intensity (see inset Fig. 4). A two-phase refinement with both the orthorhombic and monoclinic space groups is found to be acceptable, and all the observed diffraction peaks could be indexed (see Fig. 4 bottom). We find that the orthorhombic and monoclinic phases coexist over a large temperature range from 150 to 6 K in cooling cycle and from 6 to 240 K in heating cycle. The percentage of the orthorhombic phase continuously decreases with lowering temperature and reaches 26 % at 6 K for x=0.03 and becomes zero for x=0.05, respectively. The coexistence of both the phases indicates that the structural phase transition from orthorhombic to monoclinic phase is of first order for both compounds. Structural parameters obtained after Rietveld refinement for both compounds are given in table S2 (supplementary materials). The observed coexisting phases and anomalous smearing of the dielectric response is akin to dipolar glasses and relaxor ferroelectrics. The observed monoclinic phase (Cc) is found to be identical to PZT where piezo response is ultra large. This monoclinic phase is believed to provide for a flexible polarization rotation and considered to be directly linked to the high performance piezoelectricity in materials due to presence of more easy axis for spontaneous polarizations.

The variation of lattice parameters obtained from the Rietveld refinements for composition x=0.03 and 0.05 with temperature are plotted in Fig. 5 and 6. It is evident from the figure that for composition x=0.03, in the cooling cycle, the $a_o$ and $c_o$ lattice parameters of the orthorhombic phase decrease and show an anomaly around 120 K. In the heating cycle, the lattice parameters for both phases (orthorhombic and monoclinic) increase and show an anomaly around 250 K. These are the temperature, where we see an anomaly in dielectric studies of the compound. These small anomalous jumps of less that about 0.05 % in lattice constants may reflect release of some strain in the sample that builds up during the process of the phase transition. A small hysteresis in the lattice constants of the same order of 0.05 % also indicates that the relaxation process is much slower than the time scale of several hours in the experiment. It is also important to notice that the monoclinic phase has a higher volume compared with the orthorhombic phase. However, for the composition x=0.05, the lattice parameters decrease on decreasing temperature and increases on increasing temperature. The variation of phase fraction with temperature suggests that the monoclinic phase is dominant below 75 K for both the compounds. It is also clear that the composition x=0.05 completely transforms into monoclinic phase and becomes single phase below 50 K. On the other hand, the composition x=0.03 does not completely transform into monoclinic phase even at the lowest temperature (6 K). The thermal hysteresis for composition x=0.03 is larger than that of x=0.05. This suggests that addition of increasing amounts of $BaTiO_3$ in $NaNbO_3$ suppresses the thermal hysteresis. The structural phase transition temperature from orthorhombic to monoclinic phase decreases on increasing concentration of $BaTiO_3$. Variation of lattice parameter for the orthorhombic phase for composition x=0.03 also exhibits a thermal hysteresis which can be signature for ferroelastic type phase transition. [39-41] It is well documented in literature that strain plays a major role in structural phase transition in perovskite structure and which is responsible for ferroelastic phase transition[39-41]. In the present case, the composition x=0.03 exhibits phase coexistence and the low-temperature monoclinic phase has a higher volume compared to the high-temperature orthorhombic phase. Hence as the monoclinic phase grows, it creates strain in the system.

In conclusion, we have carried out systematic temperature-dependent neutron diffraction measurements in conjunction with dielectric spectroscopy from 6 to 300 K. The temperature dependent dielectric constant measurements show dispersion with frequency, and exhibit anomalies at different temperatures in cooling and heating cycles. In order to explore structural change as a function of temperature, we analyzed the powder neutron diffraction data collected in the same temperature range. The disappearance of certain superlattice reflections and appearance of splitting in main cubic-perovskite peaks at low temperatures provide a signature of structural phase transition. Detailed Rietveld analysis of powder neutron diffraction data reveals that the ferroelectric orthorhombic and rhombohedral structural models are unsatisfactory. We have explored various possibilities and found that the monoclinic phase with space group Cc accounted nearly all the reflections at lowest temperature. This monoclinic phase is believed to provide for a flexible polarization rotation and considered to be directly linked to the high performance piezoelectricity in materials. The coexistence of both the phases indicates that the structural phase transition from orthorhombic to monoclinic phase is of first order. The coexisting phases and reported anomalous smearing of the dielectric response are akin to dipolar glasses and relaxor ferroelectrics.

See Supplementary Materials for the details of sample perpetration, experiment, data analysis, results of Rietveld refinements, and results of temperature dependent PE loop.

S. L. Chaplot would like to thank the Department of Atomic Energy, India for the award of Raja Ramanna Fellowship.

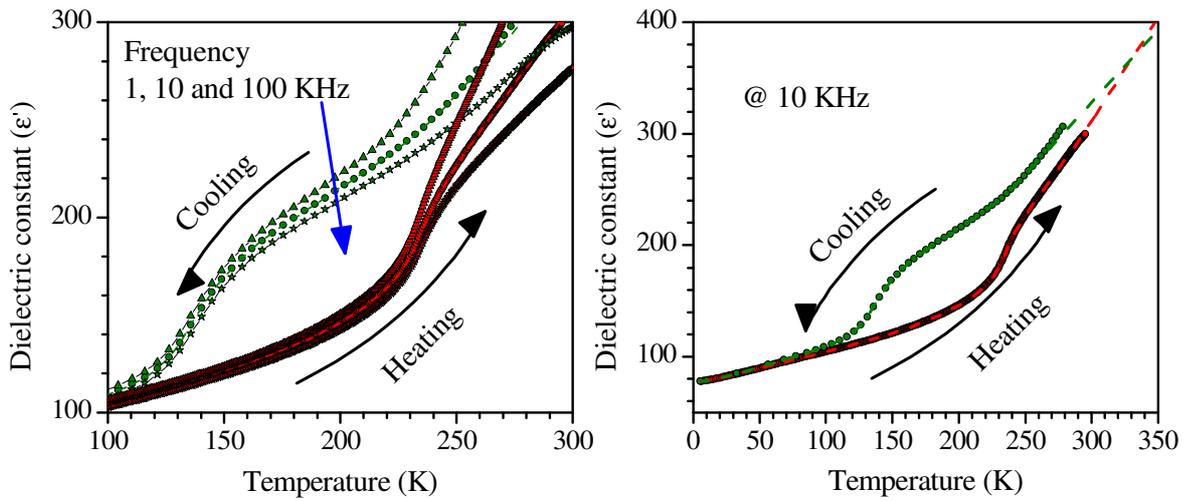

**Figure 1.** Temperature variation of dielectric constant of 0.97 NaNbO$_3$ -0.03BaTiO$_3$ at selected frequencies.

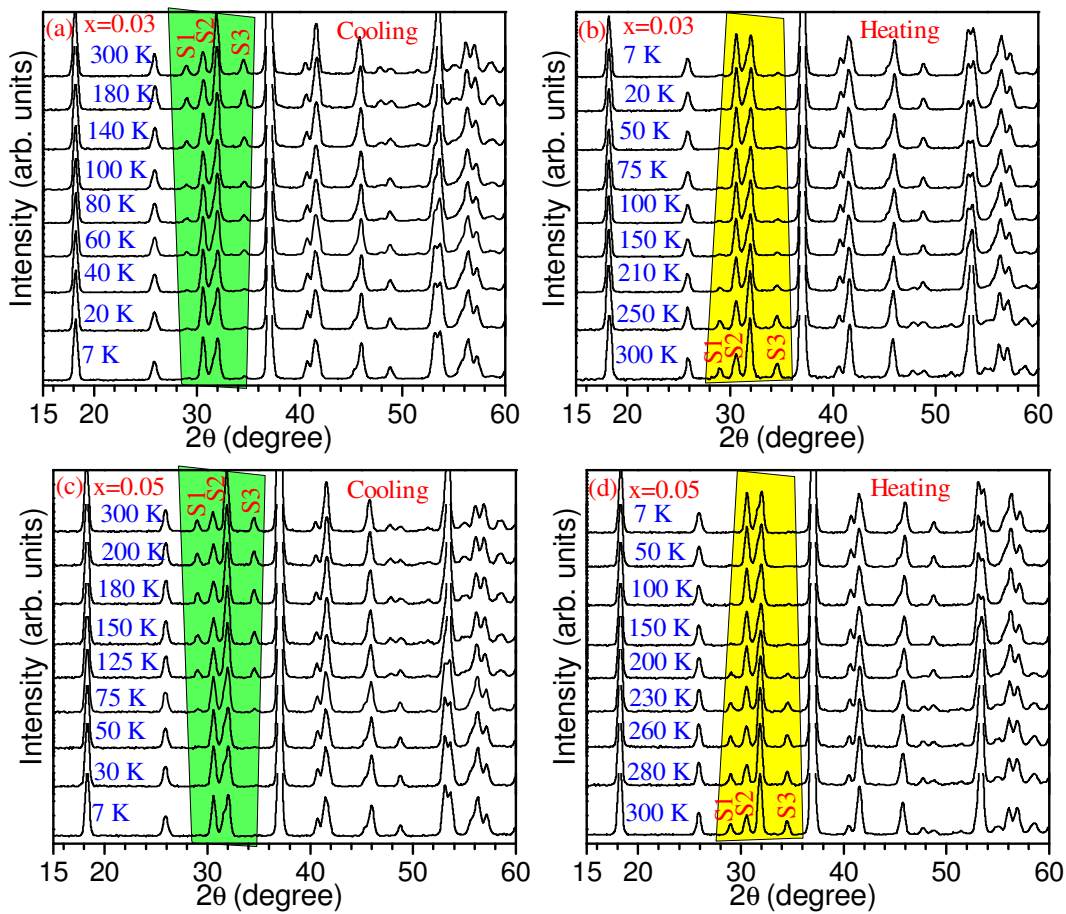

**Figure 2.** Evolution of the neutron diffraction pattern for (1-x) NaNbO$_3$-x BaTiO$_3$ for (a, b) x=0.03, (c, d) and 0.05 with temperature.

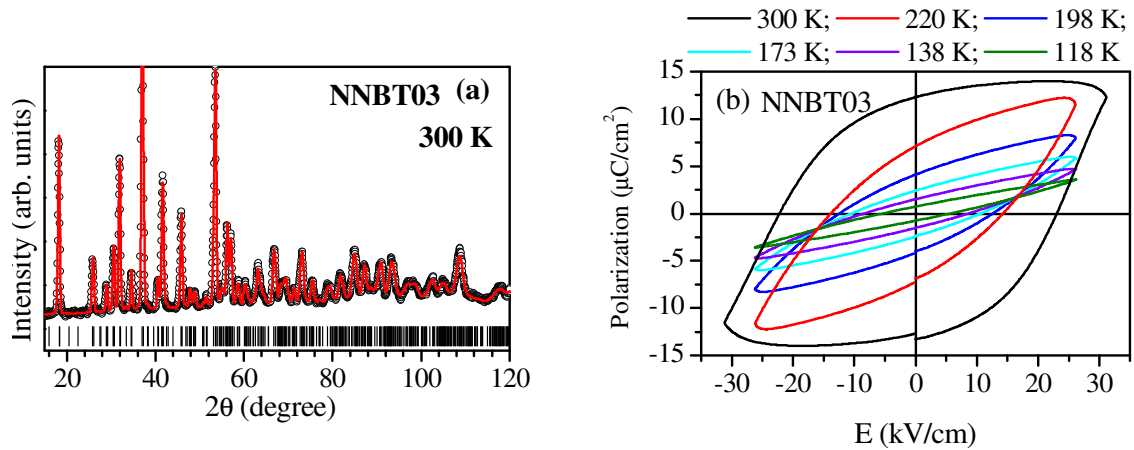

**Figure 3**. (a) Observed (dot), and calculated (continuous line) profiles obtained after the Rietveld refinement using orthorhombic ($Pmc2_1$) for x=0.03. (b) Polarization vs. Electric Field (P-E) curve for NNBT03 at selected low temperatures

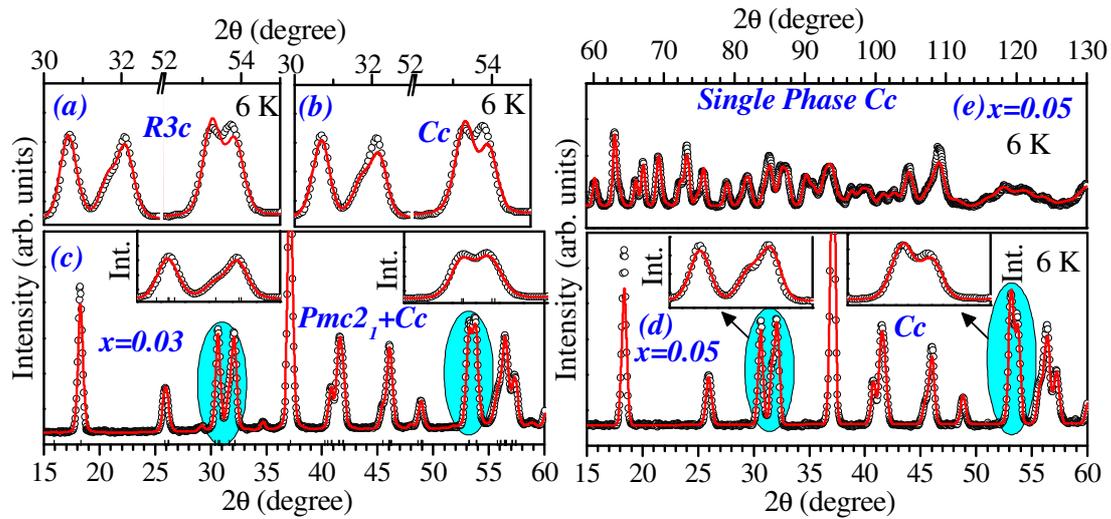

**Figure 4**. Observed (dot) and calculated (continuous line) profiles obtained after the Rietveld refinement using (a) rhombohedral ($R3c$), (b) monoclinic ($Cc$), (c) using both orthorhombic ($Pmc2_1$) and monoclinic phases ($Cc$) for x=0.03, and (d and e) single phase monoclinic phase (Cc) for x=0.05 at 6 K. The insets of (c and d) show quality of fitting in highlighted regions using phase coexistence for x=0.03 and single phase for x=0.05 respectively. In all the refinements, the data over full angular range has been used, although in the figures only a limited range is shown to highlight the changes clearly.

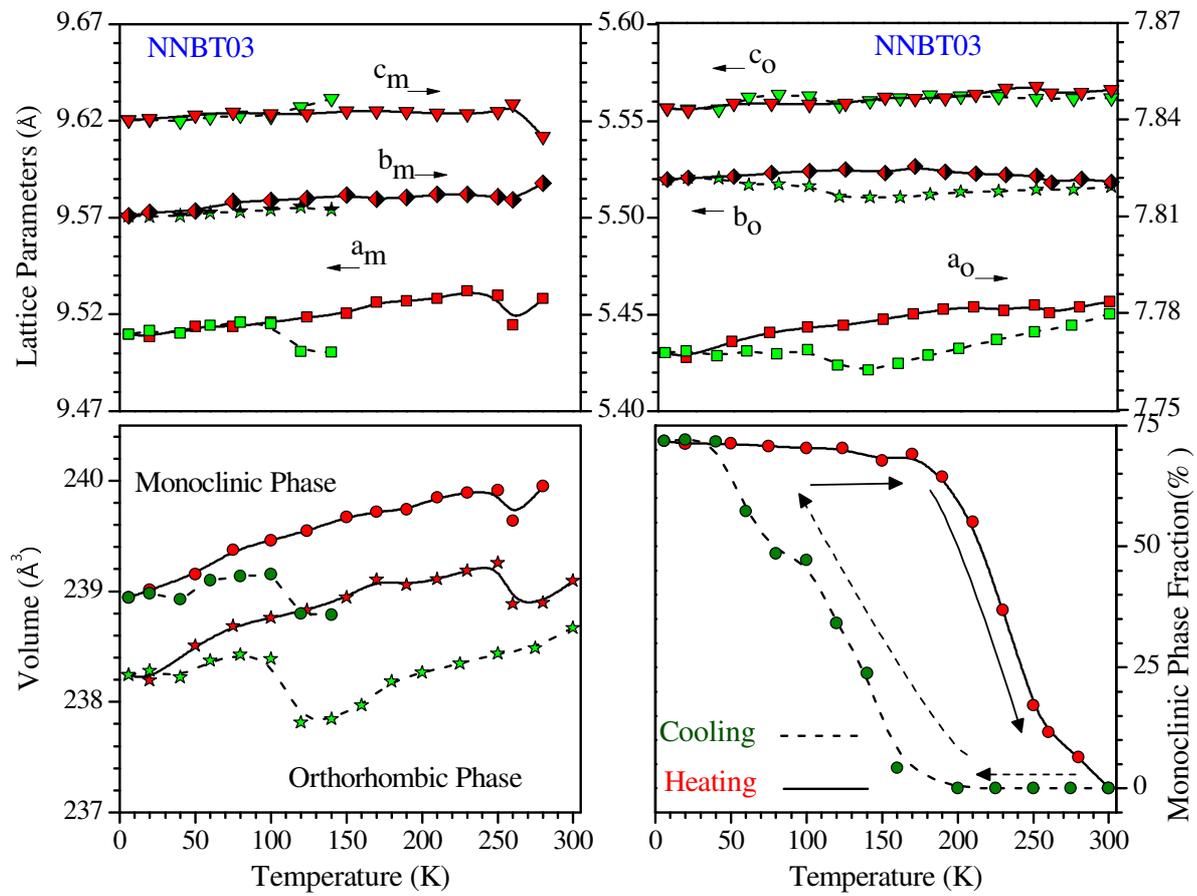

**Figure 5.** Evolution of structural parameters obtained after Rietveld analysis of powder neutron diffraction data of the composition NNBT03 with temperature. Solid and dash lines correspond to heating and cooling cycles respectively.

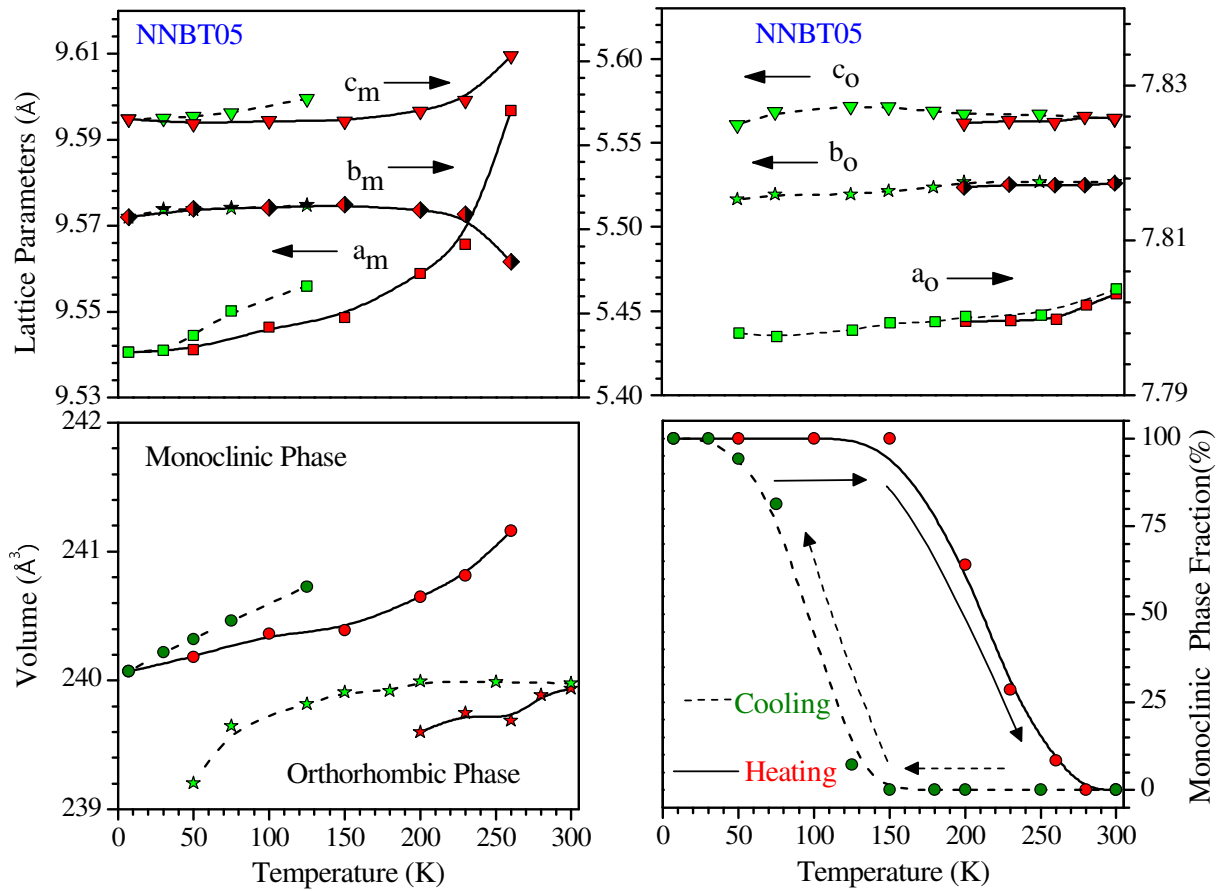

**Figure 6.** Evolution of structural parameters obtained after Rietveld analysis of powder neutron diffraction data of the composition NNBT05 with temperature. Solid and dash lines correspond to heating and cooling cycles respectively.

# Supplementary Materials

# Evidence for existence of Functional Monoclinic Phase in Sodium Niobate based Solid Solution by Powder Neutron Diffraction

S. K. Mishra[1], Mrinal Jauhari[1,2], R. Mittal[1,2], P. S. R. Krishna[1], V. R. Reddy[3], and S. L. Chaplot[1,2]

[1]Solid State Physics Division, Bhabha Atomic Research Centre, Trombay, Mumbai- 400085, India-

[2]Homi Bhabha National Institute, Anushaktinagar, Mumbai 400094, India

[3]UGC-DAE Consortium for Scientific Research, University Campus, Khandwa Road, Indore 452001, India

Samples of (1-x) $NaNbO_3$ -x$BaTiO_3$(x=0.03 and 0.05) were prepared by solid state reaction method. The single phase purity was confirmed by powder X-ray diffraction. The neutron powder diffraction data have been recorded on sample in the 2θ range of 4°–138° with a step of width of 0.1° using neutrons of wavelength of 1.2443 Å on a medium resolution position sensitive detector based powder diffractometer at the Dhruva Reactor in Bhabha Atomic Research Centre. An 8 mm diameter thin-walled vanadium can of length 55 mm was used to hold the powder sample. The temperature dependent measurements from 6 K to 300 K were carried out using a closed cycle refrigerator. The accuracy of the measured temperature during data collection was within ±0.2 K. All the diffraction measurements have been carried out in both cooling and heating cycles. The structural refinements were performed using the Rietveld refinement program FULLPROF.

Low-temperature dielectric measurements were carried out using a frequency-response analyzer (Novocontrol TB-Analyzer). For cooling the sample down to 5 K, a closed-cycle refrigerator with He-gas exchange attachment was used. Temperature-dependent dielectric constant was measured in the heating cycle between 5 to 300 K and in the frequency range of 100 Hz to 1 MHz with a heating rate of 0.8 K/min. The polarization versus electric field (P-E) hysteresis loops at selected temperature were studied using a ferroelectric loop (P-E) tracer of M/s Radiant Instruments, USA.

**Table S1:** Structural parameters of NNBTx with x=0.03 and 0.05 obtained by Rietveld refinement of powder neutron diffraction using orthorhombic structure (space group $Pmc2_1$) at room temperature.

| Atoms | Site | Composition x=0.03 at T= 300 K | | | | Composition x=0.05 at T= 300 K | | | |
|---|---|---|---|---|---|---|---|---|---|
| | | Positional Coordinates | | | | Positional Coordinates | | | |
| | | x | y | z | B (Å$^2$) | x | y | z | B (Å$^2$) |
| Na1/Ba1 | 2a | 0.000 | 0.2523(3) | 0.2414(1) | 0.30(2) | 0.000 | 0.2649(2) | 0.2991(2) | 0.25(2) |
| Na2/Ba2 | 2b | 0.500 | 0.7690(3) | 0.7632(1) | 0.90(2 | 0.500 | 0.7725(2) | 0.8013(2) | 0.55(3) |
| Nb/Ti | 4c | 0.7484(4) | 0.7463(1) | 0.2427(3) | 0.12(6) | 0.7497(5) | 0.7504(1) | 0.3261(2) | 0.21(3) |
| O1 | 2a | 0.00 | 0.1978(3) | 0.7206(5) | 0.64(4) | 0.00 | 0.1947(5) | 0.800(1) | 0.84(3) |
| O2 | 2b | 0.500 | 0.7013(3) | 0.2051(4) | 0.62(1) | 0.500 | 0.7088(4) | 0.2919(6) | 1.39(1) |
| O3 | 4c | 0.2231(9) | 0.4576(3) | 0.4355(3) | 0.63(3) | 0.2308(7) | 0.4637(3) | 0.5112(3) | 0.54(8) |
| O4 | 4c | 0.2744(1) | 0.0250(1) | -0.008(2) | 0.60(9) | 0.2780(4) | 0.0310(1) | 0.0759(2) | 0.80(4) |
| $a_o$= 7.7797(4) (Å); $b_o$= 5.5159 (3) (Å); $c_o$= 5.5618(2) (Å), unit-cell volume 238.66(2) (Å$^3$) $R_p$= 2.79; $R_{wp}$= 3.82; $R_{exp}$=3.40; $\chi^2$= 1.26 |||||| $a_o$= 7.8031(2) (Å); $b_o$= 5.5259 (2) (Å); $c_o$= 5.5646 (2) (Å), unit-cell volume= 239.94(1) (Å$^3$) $R_p$= 3.24; $R_{wp}$= 4.38 $R_{exp}$=3.12; $\chi^2$= 1.97 ||||

**Table S2:** Structural parameters of NNBTx with x=0.03 obtained by Rietveld refinement of powder neutron diffraction data at 6 K using two phase model monoclinic phase (space group Cc) and orthorhombic (space group: $Pmc2_1$).

| Atoms | Composition x=0.03 at T= 6 K Space group *Cc* (phase fraction 74%) | | | | | Composition x=0.03 at T= 6 K Space group *Pmc2$_1$* (phase fraction 26 %) | | | | |
|---|---|---|---|---|---|---|---|---|---|---|
| | Site | Positional Coordinates | | | | Site | Positional Coordinates | | | |
| | | x | y | z | B (Å$^2$) | | x | y | z | B (Å$^2$) |
| Na1/Ba1 | 4a | 0.000 | 0.7500 | 0.0000 | 0.07(3) | 2a | 0.000 | 0.240(1) | 0.269(3) | 0.03(3) |
| Na2/Ba2 | | | | | | 2b | 0.500 | 0.7684(4) | 0.7854(2) | 0.04(1) |
| Nb/Ti | 4a | 0.2640(1) | 0.2506(3) | 0.2699(2) | 0.04(5) | 4c | 0.7506(7) | 0.7533(5) | 0.2477(3) | 0.04(2) |
| O1 | 4a | 0.0265(3) | 0.3036(3) | 0.0675(3) | 0.18(6) | 2a | 0.0000 | 0.216(3) | 0.7304(6) | 0.50(2) |
| O2 | 4a | 0.3102(2) | 0.4664(4) | 0.0600(2) | 0.02(2) | 2b | 0.5000 | 0.260(3) | 0.3213(6) | 0.26(4) |
| O3 | 4a | 0.2472(3) | 0.0302(3) | 0.5698(6) | 0.06(3) | 4c | 0.225(3) | 0.494(4) | 0.109(6) | 0.38(3) |
| O4 | | | | | | 4c | 0.287(1) | 0.0152(3) | 0.000(3) | 0.08(2) |
| $a_m$= 9.5054(4) (Å); $b_m$= 5.5012 (2) (Å); $c_m$= 5.5512 (2) (Å), Beta angle= 124.63 (1); unite cell volume= 238.94 (2) (Å$^3$) |||||| $a_o$= 7.7681(1) (Å); $b_o$= 5.5191 (2) (Å); $c_o$= 5.5571 (1) (Å), unit cell volume= 238.25 (5) (Å$^3$) ||||
| $R_p$= 2.78; $R_{wp}$= 3.73; $R_{exp}$=3.39; $\chi^2$= 1.21 ||||||||||